\documentclass[preprint,pre,showpacs,preprintnumbers,amsmath,amssymb,unsortedaddress]{revtex4}

\usepackage{hyperref}%
\usepackage{graphicx}
\usepackage{dcolumn}
\usepackage{bm}
\usepackage{epsf}
\usepackage{mathrsfs}

\usepackage{epsf}

\begin{document}

\title{Thermomechanical effects in uniformly aligned
dye-doped nematic liquid crystals}

\author{Dmitry O. Krimer$^{(1)}$ and
Stefania Residori$^{(2)}$}
\affiliation{$^{(1)}$ Theoretische
Physik, Universitaet Tuebingen, 72076 Tuebingen, Germany
\\
$^{(2)}$ Institut Non Lin\'eaire de Nice, 1361 route des Lucioles,
 06560 Valbonne, France}

\date{\today}

\begin{abstract}

We show theoretically that thermomechanical effects in dye-doped
nematic liquid crystals when illuminated by laser beams, can
become important and lead to molecular reorientation at
intensities substantially lower than that needed for optical
Fr\'eedericksz transition. We propose a 1D model that assumes
homogenous intensity distribution in the plane of the layer and is
capable to describe such a thermally induced threshold lowering.
We consider a particular geometry, with a linearly polarized light
incident perpendicularly on a layer of homeotropically aligned
dye-doped nematics.
\end{abstract}

\pacs{05.45.-a, 42.70.Df, 42.65.Sf}
\maketitle

\section{Introduction}

Optically induced temperature changes in liquid crystals are at
the origin of interesting nonlinear behaviors \cite{Simoni}.
Indeed, due to the light absorption, thermal effects can change
the physical properties of the liquid crystal, which in turn
affects the light propagation in the medium. Among the examples of
thermal effects, we can distinguish between a direct change of the
refractive indices, often referred to as thermal indexing, and the
variation with the temperature of other physical parameters, such
as the elastic constants, which  may influence the light
propagation as well. In the latter case, thermal effects might be
responsible for director reorientation, and are  thus referred to
as thermomechanical effects.

Previously, light-induced thermomechanical effects have been
largely investigated in cholesteric liquid crystals, where these
effects were originally related with the absence of the right-left
symmetry \cite{Leslie}. Then it became clear that such effects
might also exist in systems which possesses this symmetry, such as
the nematic liquid crystals (NLC). As was shown in
\cite{Akop_84,Brand_87}, thermomechanical effects do give a
contribution to the director, heat and Navier-Stokes equations.
These thermomechanical terms are given by nonlinear combinations
with respect to temperature, director and velocity gradients and
represent nonlinear cross couplings between them. The first
question which arose is to suggest an experiment for measuring the
magnitude of thermomechanical coefficients. Indeed, it is hardly
possible to find a setup which leads to the contribution of one of
the coefficients only.  Since for uniformly aligned nematics no
thermomechanical effects are expected, the hybrid-oriented
nematics were used in the experiments \cite{Akop_97,Akop_01}.
There, the hydrodynamic flow appeared as a result of the applied
temperature gradient, which allowed to find the magnitude for the
thermomechanical coefficients. Laser induced thermomechanical
effects in dye-doped nematics have been envisaged in preliminary
experiments \cite{Barnik}. However, it was not possible to derive
a definitive conclusion due to the main difficulty of separating
thermomechanical contributions from the light-induced molecular
torque. Another problem when performing these experiments is that
of avoiding to approach the nematic-isotropic transition, where an
enhancement of the nonlinear optical response of dye-doped
nematics could take place because of other effects, such as the
weakening of the anchoring \cite{Simoni2}.

Here, we study theoretically the thermomechanical effects which
occur in uniformly aligned dye-doped nematic liquid crystals. This
happens when the sample is illuminated by a laser beam with a
wavelength in the absorption band of the dye, which causes a
significant heating of the liquid crystal layer. Note that the
absorbtion is negligible for pure nematics, so that
thermomechanical effects are significant only for dye-doped
nematics. Indeed, in the presence of dye-doping thermal heating
leads to an additional torque which acts onto the director
together with the light-induced torque. This additional torque
will help either to destabilize or stabilize the initial
orientation which is determined by the ratio of the
thermomechanical coefficients. We show that the additional torque
might be strong enough to induce the molecular reorientation well
before the onset of the light-induced molecular reorientation, the
so called optical Fr\'eedericksz transition (OFT) for pure liquid
crystals \cite{Zeldovich} and Janossy effect for dye-doped liquid
crystals \cite{JAN_90,Jan-91}. We account for the possible
decrease of the reorientation threshold by a 1D model that assumes
homogenous intensity distribution in the plane of the layer (the
plane wave approximation) and includes the light absorption into
the hydrodynamic equations for the nematics.

\section{Theoretical model}

In dye-doped nematics, contrary to the pure ones, the OFT lowering
might happen not only because of the well known Janossy effect
\cite{JAN_90,Jan-91} but also because thermomechanical effects may
add a significant contribution. Indeed, as a consequence of light
absorption, light propagation in dye-doped nematics causes
significant heating of the LC. When the intensity of a beam is
sufficiently large and temperature gradient becomes nonzero, an
additional force acting on the director and additional terms in
the stress tensor in Navier-Stokes equation for the velocity
appear due to thermomechanical effect \cite{Akop_84, Brand_87}.
The equation for the velocity is coupled with the director
equation, so any dynamical process that leads to director
reorientation will also induce flow even in the absence of
pressure gradients. It should be noted that the terms which
describe the thermomechanical effects appear not only in the
director but also in Navier-Stokes equations. Thus, neglecting the
velocity equations and considering only the director equation may
lead to misleading results, even though this simplification has
been done for many studies in the context of light induced
instabilities.

\begin{figure}[h!]
\center {\resizebox{8cm}{!}{\includegraphics{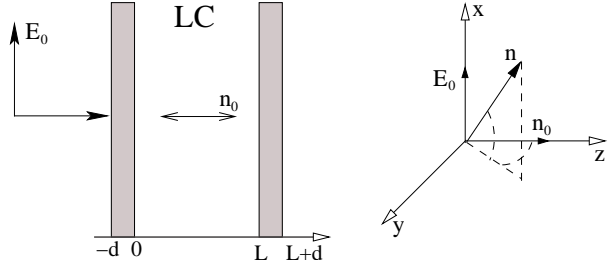}}}
\caption{\label{fig_geom} Geometry of the setup: a linearly
polarized light along the ${\bf x}$-direction is at a normal
incidence on a nematic LC layer;  ${\bf n_0}
\parallel {\bf z} $ is the director of the initial unperturbed alignment
(homeotropic state).}
\end{figure}

In our case, we considered the full problem with the velocity
equations coupled to the director reorientation equation. However,
in order to calculate the shift of the OFT threshold due to the
thermomechanical effect we employed another frequently used
simplification, namely that all variables depend only on one
coordinate, which is transversal to the plane of the nematic layer
(1D assumption). We thus considered a linearly-polarized plane
wave incident perpendicularly on a layer of a dye-doped nematic of
thickness $L$ that is sandwiched between two substrates of
thickness $d$ (see Fig.~\ref{fig_geom}). The cell has initially
homeotropic alignment (with strong homeotropic anchoring at the
boundaries) and is placed in a thermostage with a temperature
$T_0$ on both sides. The light is polarized along the ${\bf
x}$-direction and propagates along the positive  ${\bf z}$-axis.
\subsection{Heat equation}

For simplicity we also assumed that i) the attenuation of the
light inside the nematic is small [i.e. $I(z)\simeq I_0$, where
$I_0$ is the incident intensity]; ii) the transversal heat flow
occurring in the $({\bf x,y})$ plane is neglected, i.e. the
temperature profile depends only on $z$, $T=T(z)$. Since we deal
with the plane-wave approximation the following 1D steady-state
heat conductivity equations in the nematic and substrates can be
written (note that an absorption takes place only inside the
nematic):
\begin{eqnarray}
\label{heat_eqs} &&\kappa_s\partial_z^2T=0,
\,\,\,\,-d\le z\le 0 \,\,\text{or} \,\, L\le z\le L+d
\\\nonumber
&&\kappa_{||}\partial_z^2T=-\alpha_{\perp}I_0,\,\,\,\, 0\le z\le L \,,
\end{eqnarray}
where $\kappa_{||}$ is the parallel component of the heat
conductivity tensor of the nematic, $\kappa_s$ is the heat
conductivity of the substrates and $\alpha_{\perp}$ is the
absorption coefficient for the ordinary light. We then write the
boundary conditions given by continuity conditions of the
temperature and the heat flow at the substrate-nematic interfaces
($z=0$ and $z=L$):
\begin{eqnarray}
\label{bound_c_T} &&T_{s1}|_{z=0}=T_N|_{z=0},
\,\,\kappa_s\partial_zT_{s1}|_{z=0}=\kappa_{||}\partial_zT_N|_{z=0}
\\
&&T_N|_{z=L}=T_{s2}|_{z=L},\,\,
\kappa_{||}\partial_zT_N|_{z=L}=\kappa_s\partial_zT_{s2}|_{z=L}.
\end{eqnarray}
In this simplest model the temperature profile is linear inside
the substrates and has a parabolic form inside the nematic which
is symmetric with respect to the center of the layer $z=L/2$. The
maximal temperature is at the center of the layer and the maximal
temperature difference inside the nematic is $\Delta
T_{max}=\alpha_{\perp}I_0 L^2/(8 \kappa_{||})$. The temperature
gradient inside the nematic can be written as
\begin{eqnarray}
\label{TN_grad}
\partial_z T_N = \beta \, \rho \,(L-2z),\,\, \text{with}\,\,
\beta=\dfrac{\alpha_{\perp} I_F}{2 \kappa_{||}},
\end{eqnarray}
where $\rho=I_0/I_{F}$ is the incident intensity normalized to the
threshold intensity of the OFT for a dye-doped nematic \cite{Janossy-99}
\begin{eqnarray}
\label{porog} I_{F}=\dfrac{\pi^2}{L^2}
\dfrac{c (\varepsilon_\perp+\varepsilon_a)K_3}
 {\varepsilon_a\sqrt{\varepsilon_\perp}
\eta},{~}\eta = \dfrac{\varepsilon_a+\zeta}{\varepsilon_a}.
\end{eqnarray}
Here $\varepsilon_a=\varepsilon_\|-\varepsilon_\perp$ is the
dielectric anisotropy and $\varepsilon_\perp
{~}$($\varepsilon_\|$) is the dielectric permittivity (at optical
frequency) perpendicular (parallel) to ${\bf n}$, $\zeta$
phenomenologically describes the effect of certain dye dopants
($\zeta=0$ in a pure LC), $K_3$ is the bend elastic constant of
the nematic and $c$ is the velocity of light in the vacuum.

It should be noted that the obtained solution (\ref{TN_grad}) for
the temperature profile inside the nematics is much simpler than
in reality. The more complicated solution for Gaussian incident
beams has been derived in \cite{JAN_temp_91} from 2D heat equation
which includes transversal dependence. It turned out that for
large Gaussian beams i.e. when the spot size $w$ is much larger
than the thickness of the layer $L$, $w \gg L$, the maximum
temperature rise, $\Delta T_{max}$, becomes proportional to the
spot size. Such a behavior is not predicted by the 1D model and is
owing to the transversal heat flow occurring in the plane of the
layer. Nevertheless, the realization of the plane wave
approximation in the experiment is difficult but not impossible
task. One of the way to proceed is to enlarge the Gaussian beam to
a size which is much larger than the medium working area (which is
much stronger condition then $w \gg L$), so that the intensity can
be considered uniform in the central part. Another possibility is
that of using the so-called flat-top beams \cite{flat-top}. Both
techniques are quite easily accessible and allow to attain the
threshold for OFT. Preliminary experiments are running at present
in our laboratory and will be reported elsewhere.

\subsection{Linearized Navier-Stokes equation}

The Navier-Stokes equation for the velocity ${\bf v}$ can be
written as \cite{de_Gennes}
\begin{eqnarray}
\label{bas_Nav_St} \rho_m \, (\partial_t+{\bf v}\cdot\nabla)
v_i=-\nabla_j\left(p\,\delta_{ij}+\pi_{ij}+
T_{ij}^{visc}+T_{ij}^{TM}\right),
\end{eqnarray}
where $\rho_m$ and $p$ are the density and the pressure of the LC,
respectively.  $\pi_{ij}$ is the Ericksen stress tensor
\cite{de_Gennes}. The viscous stress tensor $T_{ij}^{visc}$ in
Eq.~(\ref{bas_Nav_St}) is written in terms of the six Leslie
coefficients $\alpha_i$\cite{de_Gennes} and the thermomechanical
tensor $T_{ij}^{TM}$ is introduced in \cite{Akop_84,Brand_87}.
Then, the incompressibility condition ($\rho_m$ is constant)
$\nabla\cdot{\bf v}=0$ and the no-slip boundary conditions ${\bf
v}|_{z=0,L}=0$ immediately ensure that the $z$ component of the
velocity vanishes ${\bf v}=(v_x(z,t),v_y(z,t),0)$, so ${\bf v}$ is
parallel to the plane of the layer. Moreover, all convective
derivatives  ${\bf v} \cdot \nabla$ vanish. The Navier-Stokes
equation has been simplified then due to the fact that the
director relaxation time
\begin{eqnarray}
\label{tau_backfl} \tau=\dfrac{\gamma_1 L^2}{\pi^2 K_3}
\end{eqnarray}
differs by many orders of magnitude with the momentum diffusion
time $\tau_{visc}=\rho_mL^2/\gamma_1$, where
$\gamma_1=\alpha_3-\alpha_2$ is the rotational viscosity.
Typically $\tau_{visc}\sim 10^{-6}\,s$ and $\tau\sim 1 s$, so the
slow variable of the system is the evolution of the director which
enslaves the flow motion and, thus, the inertial terms in the
Eq.~(\ref{bas_Nav_St}) can be neglected. Taking into account that
the light is polarized in the ${\bf x}$-direction, we need an
equation for the $x$ component only. The linearization of this
equation around the homeotropic state ($n_x=0,\,v_x=0$) yields
\begin{eqnarray}
\label{eq_1_NS}
&&(\alpha_5-\alpha_2+\alpha_4) \dfrac{\partial_z
v_x}{2}+\alpha_2
\partial_t n_x-
\\\nonumber
&&a_{10}\partial_z n_x\partial_z T_N=C(t),\:
\end{eqnarray}
where $C(t)$ is a function that does not depend on $z$ and will be
fixed by the boundary conditions. Here the first two terms on the
left-hand side of Eq.~(\ref{eq_1_NS}) come from the viscous stress
tensor whereas the last one is the contribution from the
thermomechanical stress tensor with the thermomechanical
coefficient $a_{10}$. (Note that $a_{10}$ in \cite{Brand_87} is
related to the $\xi_i$ from \cite{Akop_84} as
$a_{10}=(\xi_8-\xi_4)/4$.) The unknown function $C(t)$ can be
determined by integrating Eq.~(\ref{eq_1_NS}) across the layer.
Finally, the velocity gradient $\partial_z v_x$ can be expressed
in terms of the director and the temperature gradient [which is
given by Eq.~(\ref{TN_grad})] as:
\begin{eqnarray}
\label{grad_vy_haupt} &&\partial_z v_x=
\dfrac{2}{\alpha_5-\alpha_2+\alpha_4}\times
\\\nonumber
&&\left[C(t)-
 \alpha_2\partial_tn_x+a_{10} \beta
\rho (L-2z) \:\partial_z n_x \right]\,
\\
&&C(t)=\dfrac{1}{L}\left(\alpha_2 \int \limits_0^{L} \: \partial_t
n_x\,dz- 2 a_{10} \:\beta \rho \int \limits_0^{L} \: n_x
\,dz\right).\;\;\;\;
\end{eqnarray}
%

\subsection{Linearized director equation.
Adiabatic elimination of the flow field}

The equation for the director ${\bf n}$ is
\begin{eqnarray}
\gamma_1(\partial_t +{\bf v}\cdot\nabla-\bm{\omega}\times)
\:{\bf n} =-\underline{\underline\delta}^{\perp}\,
 (\gamma_2\underline{\underline A}{\bf n}+{\bf h}-{\bf g}^{TM})\,,
\label{bas_dir_with_flow}
\end{eqnarray}
where $\gamma_2=\alpha_3+\alpha_2$. ${\bf h}$ is the molecular
field obtained from the variational derivatives of the free energy
density $F$, which consists of the elastic and the electrical
parts \cite{de_Gennes}. The projection operator
$\delta_{ij}^{\perp}=\delta_{ij}-n_i n_j$ in
Eq.~(\ref{bas_dir_with_flow}) ensures conservation of the
normalization ${\bf n}^2=1$. Here $A_{ij}$ is the symmetric
strain-rate tensor and the vector ${\bf N}$ gives the rate of
change of the director relative to the fluid. In
Eq.~(\ref{bas_dir_with_flow}) ${\bf g}^{TM}$ is the contribution
of the thermomechanical effect to the force acting on the director
\cite{Akop_84}.

We then linearized Eq.~(\ref{bas_dir_with_flow}) around the
homeotropic state and obtained the following equation for $n_x$
\begin{eqnarray}
\label{lin_dir_1} &&\gamma_1\partial_t n_x+\alpha_2 \partial_z
v_x=
\\\nonumber
&&K_3 \left[\partial_z^2 n_x+\left(\dfrac{\pi}{L}\right)^2\rho\:
n_x\right]- \dfrac{\xi_4}{2}\:\partial_z n_x \cdot \partial_z T_N.
\end{eqnarray}
Note that the first term in the square brackets on the right-hand
side of Eq.~(\ref{lin_dir_1}) stems from the linearization of
${\bf h}$  whereas the last one after the linearization of ${\bf
g}^{TM}$.

From here on we will use normalized time $t\rightarrow t/\tau$
[where $\tau$ is the director relaxation time, see
Eq.~(\ref{tau_backfl})], length $z\rightarrow \pi z/L$ (the same
symbols will be kept). We will also introduce dimensionless
viscosity coefficients  $\alpha_i^{'} = \alpha_i/\gamma_1$.
Eliminating the velocity gradient from Eq.~(\ref{lin_dir_1}) with
the help of Eq.~(\ref{grad_vy_haupt}) and using the expression for
the temperature gradient (\ref{TN_grad}), the following equation
for $n_x$ can be derived
\begin{eqnarray}
\label{lin_dir_haupt} &&\partial_z^2 n_x+\rho\: n_x-
(1-b)\partial_t n_x-d_1\rho(\pi-2z)\partial_z n_x-\;\;\;\;
\\\nonumber
&& \dfrac{b}{\pi}\left\{ \int \limits_0^{\pi}
 \: \partial_t n_x\,dz - \dfrac{2d_2\rho}{\alpha_2^{'}}
 \int \limits_0^{\pi}\;
n_x\,dz \right\}=0,
\end{eqnarray}
where $b$, $d_1$ and $d_2$ are dimensionless parameters defined as follows:
\begin{eqnarray}
\label{deltas} b=\dfrac{2\alpha_2^{'2}}{\alpha_5^{'}-
\alpha_2^{'}+\alpha_4^{'}},\, d_1=\delta_1\psi,\,d_2=\delta_2
\psi.
\end{eqnarray}
Here $\psi$ depends on the absorption and $\delta_{1,2}$ are the
algebraic combinations of the thermomechanical coefficients given
by
\begin{eqnarray}
\psi&=&\dfrac{\beta\tau}{\gamma_1}=\dfrac{\alpha_{\perp} c
(\varepsilon_\perp+\varepsilon_a)} {2 \kappa_{||} \varepsilon_a
\sqrt{\varepsilon_\perp} \eta},
\\\nonumber
\delta_1&=&\left(\dfrac{b\:a_{10}}{\alpha_2^{'}}+\dfrac{\xi_4}{2}\right),\,
\delta_2=a_{10}.
\end{eqnarray}
%
%

\subsection{Linear stability analysis of the homeotropic state}

We look for solutions of Eq.~(\ref{lin_dir_haupt}) of the form
\begin{eqnarray}
n_x(z,t)=n_x(z)e^{\sigma t},
\end{eqnarray}
where $\sigma$ is the growth rate and obtain from Eqs.~(\ref{lin_dir_haupt})
\begin{eqnarray}
\label{lin_dir_last} &&\partial_z^2
n_x+\left[\rho-\sigma(1-b)\right] \: n_x-d_1\rho(\pi-2z)\partial_z
n_x-\;\;\;\;
\\\nonumber
&&\dfrac{b}{\pi}\left(\sigma-\dfrac{2d_2\rho}{\alpha_2^{'}}\right)
 \int \limits_0^{\pi} \: n_x\,dz=0\,.
\end{eqnarray}
It should be noted that Eq.~(\ref{lin_dir_last}) reduces to the
classical linearized equation for the OFT when the flow and the
thermomechanical effect are neglected by putting $b=0$ and
$d_1=d_2=0$.

Taking into account the boundary conditions $n_x|_{z=0,\pi}=0$,
Eq.~(\ref{lin_dir_last}) is solved by
\begin{widetext}
\begin{eqnarray}
\label{ny_sol} n_x=\dfrac{A}{\rho-
\sigma(1-b)}\left\{-1+e^{d_1\rho(\pi-z)z}
\dfrac{{_1}F_1\left[\dfrac{1}{2}-
\dfrac{\rho-\sigma(1-b)}{4d_1\rho},
\:\dfrac{1}{2},\:\dfrac{d_1\rho}{4}(\pi-2z)^2\right]}
{{_1}F_1\left[\dfrac{1}{2}-
\dfrac{\rho-\sigma(1-b)}{4d_1\rho},\:\dfrac{1}{2},
\:\dfrac{d_1\rho\pi^2}{4}\right]} \right\},
\end{eqnarray}
\end{widetext}
where ${_1}F_1$ is the confluent hypergeometric function
\cite{Abram} and $A$ is some constant which depends on the
director itself
\begin{eqnarray}
\label{A_def} A=-\dfrac{b}{\pi}\left(\sigma-
\dfrac{2d_2\rho}{\alpha_2^{'}}\right) \int \limits_0^{\pi} \:
n_x\,dz.
\end{eqnarray}
Substituting the solution for $n_x$ (\ref{ny_sol}) into
Eq.~(\ref{A_def}) the equation for the growth rate $\sigma$ versus
incident intensity $\rho$ can be derived (note that $A$ will be
cancelled). Substituting $\sigma=0$ into this equation, the
following transcendental equation for the critical intensity
$\rho_c$ has been obtained
\begin{align}
\nonumber
\label{rho_c} &\dfrac{\int \limits_0^{\pi}
e^{d_1\rho_c(\pi-z)z}\cdot
{_1}F_1\left[\dfrac{1}{2}-\dfrac{1}{4d_1},
\:\dfrac{1}{2},\:\dfrac{d_1\rho_c}{4}(\pi-2z)^2\right]\,dz}
{{_1}F_1\left[\dfrac{1}{2}-\dfrac{1}{4d_1},
\:\dfrac{1}{2},\:\dfrac{d_1\rho_c\pi^2}{4}\right]}=
\\
&\pi\left(1+\dfrac{\alpha_2^{'}}{2\,b\,d_2}\right).
\end{align}
\begin{figure}[h!] {
\begin{center}
\includegraphics[width=7.cm]{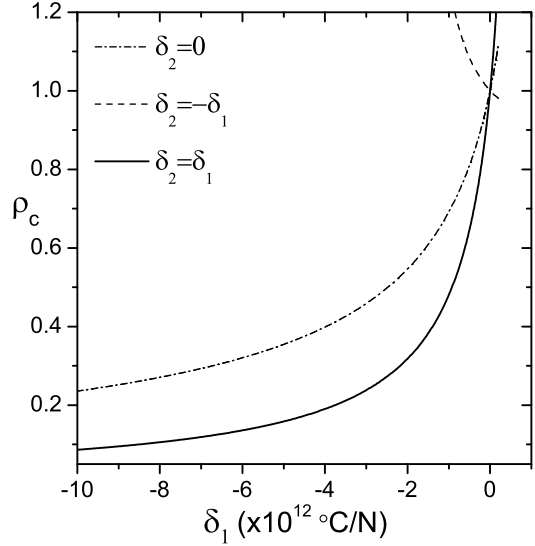}
\end{center}
\caption{The critical intensity $\rho_c$ versus parameter
$\delta_1$ under assumptions that $\delta_2=\delta_1$ (solid
line), $\delta_2=0$ (dot-dashed line), $\delta_2=-\delta_1$
(dashed line). $\rho_c=1$ corresponds to the critical intensity
for the OFT (no thermomechanical effect).} \label{fig2}}
\end{figure}
\begin{figure}[h!] {
\begin{center}
\includegraphics[width=7.cm]{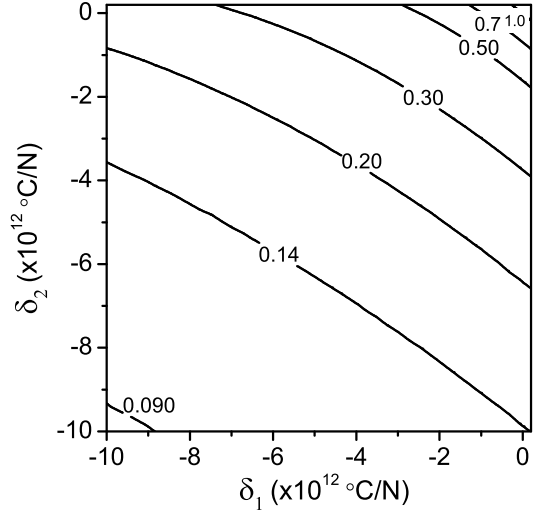}
\end{center}
\caption{Contour lines for the surface
$\rho_c(\delta_1,\delta_2)$. The values of $\rho_c$ are shown on
contour lines. ($\rho_c=1$ for $\delta_1=\delta_2=0$ corresponds
to the case without thermomechanical effect.)} \label{fig3}}
\end{figure}

In the calculations, we took the material parameters for the
nematic E7: $K_3=15.97 \times 10^{-12}$ N , $n_e=1.746$,
$n_o=1.522$ (extraordinary and ordinary refractive indices),
$\lambda=514$ $nm$, $\alpha_{\perp}=88 ~cm^{-1}$ (absorption
coefficient), $\kappa_{||}=  10^{-3} \,\text{W}/^\circ
\text{C~cm}$ (the heat conductivity for the nematic). The
calculations were made for a layer of $75~\mu$m thickness. For
these parameters and for the threshold value $I_{F}=32$ $W/cm^2$
observed in our preliminary experiment, $\zeta$ turns out to be
$\zeta \simeq 52$ which is a typical value for the enhancement
factor \cite{MAR_97} [see also Eq.~(\ref{porog})]. The maximal
temperature difference inside the layer is $\Delta T_{max}\simeq
20~K$. In the calculations we used $b=0.8$ and
$\alpha_2^{'}=-1.058$. The thermomechanical coefficients and,
hence, $\delta_1$ and $\delta_2$ are unknown. Thus we took the
typical order of their magnitudes as $10^{-12}\,N/^\circ C$
reported in \cite{Akop_01}.

In Fig.~\ref{fig2} the results of numerical solution of
Eq.~(\ref{rho_c}) [or alternatively the eigenvalue problem
(\ref{lin_dir_last})] are shown for three different cases when i)
$\delta_2=\delta_1$; ii) $\delta_2=0$; iii) $\delta_2=-\delta_1$.
($\delta_2=0$ corresponds to the thermomechanical single-constant
approximation.) One can see from this figure that $\rho_c$ might
be indeed several times lower than that for the OFT. This fact is
clearly demonstrated in Fig.~\ref{fig3} where the contour lines
for the surface $\rho_c(\delta_1,\delta_2)$ are plotted. One is
now forced to conclude that the thermomechanical effect leads
indeed to the substantial change of the critical intensity.

%
\subsection{Thermomechanical effect due to
temperature difference at the boundaries}
%

We have assumed so far that the temperature on both the bounding
plates is the same. As a next step, we have analyzed the influence
of the nonzero temperature difference $\Delta T$ maintained at the
boundaries to the instability threshold. To obtain this effect in
pure form we assumed that the thermomechanical effect is due to
$\Delta T$ only and the absorption inside nematic is neglected.
For this simple situation the temperature gradient inside nematic
is constant. Following the similar procedure described in previous
section when linearizing basic equations around the homeotropic
state [see Eq.~(\ref{lin_dir_last})], the following ODE for $n_x$
has been derived
\begin{eqnarray}
\label{lin_dir_dT} &&\partial_z^2
n_x+\left[\rho-\sigma(1-b)\right] \: n_x-d_3 \Delta T\partial_z
n_x-
\\\nonumber
&&\dfrac{b\:\sigma}{\pi} \int \limits_0^{\pi} \: n_x\,dz=0,
\end{eqnarray}
where
\begin{eqnarray}
d_3=\delta_1 \eta,\,\, \eta=\dfrac{\kappa_s L} {(\kappa_s
L+2\kappa_{||} d) \pi K_3},
\end{eqnarray}
and the other quantities have the same meaning as before. After
substituting $\sigma=0$ in Eq.~(\ref{lin_dir_dT}) the following
simple formula for the critical intensity was found
\begin{eqnarray}
\label{rho_c_dT} \rho_c=1+d_3^2 \,\dfrac{(\Delta T)^2}{4}.
\end{eqnarray}

As is seen $\rho_c$ is always higher than the threshold for OFT
and depends quadratically on $\Delta T$. This effect is, however,
small because the thermomechanical coefficient enters
quadratically to the expression for $\rho_c$ as well. (The order
of magnitude of $d_3$ is $10^{-2}\, {^\circ} C^{-1}$ for the
parameters used in calculations.)

\section{Conclusions}

In conclusion, we have shown theoretically that thermomechanical
effects might be at the origin of significant lowering of the OFT
threshold expected for dye-doped nematic liquid crystals. To
explain this, we have developed a simple model, assuming that all
physical quantities depend only on the coordinate across the
layer. We linearize both the director and Navier-Stokes equations
around the basic state, to assess the change of the primary
instability due to thermomechanical effects. The temperature
gradient across the layer, which is induced by light itself due to
absorption of the dye dopants, was calculated from the 1D heat
equation.  Using a typical value for the thermomechanical
coefficients, we have found that the effect of OFT's lowering
might be explained by thermomechanical effects. We have also
analyzed a situation when the thermomechanical effects are due to
the temperature difference maintained at the boundaries. It turned
out that in this case they always lead to an increase of the OFT
threshold.

\section{Acknowledgments}

This paper is dedicated to the memory of Professor Lorenz Kramer
who suddenly passed away  on 05/04/2005.

The authors are grateful to Dr.~Gabor Demeter for his helpful
discussions.  D.~K. gratefully acknowledges financial support by
the Deutsche Forschungsgemeinschaft under Kr 690/16.

%

%


\begin{thebibliography}{10}

\bibitem{Simoni} F. Simoni, Nonlinear Optical Properties
of Liquid Crystals and Polymer Dispersed Liquid Crystals (World
Scientific, New Jersey, 1997).

\bibitem{Leslie} F. M. Leslie, Proc. R. Soc. London, Ser. A 307, 359 (1968).

\bibitem{Akop_84}
{R.~S. Akopyan, and B.~Ya. Zel'dovich}, Sov. Phys. JETP  {\bf 60}, 953  (1984).

\bibitem{Brand_87}
{H.~R. Brand, and H. Pleiner}, Phys. Rev. A {\bf 35},  3122
(1987).

\bibitem{Akop_97}{R.~S. Akopyan,
R.~B. Alaverdyan, E.~A. Santrosyan, and Yu.~S. Chilingaryan},
Tech. Phys. Lett. {\bf 23},  690  (1997).

\bibitem{Akop_01}
{R.~S. Akopyan, R.~B. Alaverdian, E.~A. Santrosian, and Y.~S.
Chilingarian}, J. Appl. Phys. {\bf 90},  3371  (2001).

\bibitem{Barnik} M.I. Barnik, A.S. Zolot'ko,V.F. Kitaeva,
JETP {\bf 84} (6), 1122 (1997).

\bibitem{Simoni2} L. Lucchetti,M. Gentili, and F. Simoni
F.  Simoni, Appl. Phys. Lett. {\bf 86}, 151117 (2005).

\bibitem{Zeldovich} N.V. Tabirian, A.V. Sukhov,
 and B.Y. Zel'dovich, Mol. Cryst. Liq. Cryst. {\bf 136}, 1 (1986).

\bibitem{JAN_90}
I. Janossy,  A. Lloyd, and B.S. Wherrett, Mol. Cryst. Liq. Cryst.
{\bf 179}, 1  (1990).

\bibitem{Jan-91} I. Janossy, A.D. Lloyd,
Mol. Cryst. Liq. Cryst. {\bf 203}, 77 (1991).

\bibitem{Janossy-99} Istvan Janossy, J. Nonlin. Opt. Phys. Mat. {\bf 8}, 361 (1999).

\bibitem{JAN_temp_91}
I. Janossy, and T. Kosa, Mol. Cryst. Liq. Cryst. {\bf 207}, 189
(1991).

\bibitem{flat-top} J. A. Hoffnagle, and C.M. Jefferson, Appl. Opt.,
{\bf 39}, 5488 (2000).

\bibitem{de_Gennes}
P.~G. de~Gennes and J. Prost, {\em The physics of liquid crystals}
(Clarendon press, Oxford, 1993).

\bibitem{Abram}
M. Abramowitz and I.~A. Stegun, {\em Handbook of Mathematical
Functions} (Dover, New York, 1970).

\bibitem{MAR_97} L. Marrucci,
D. Paparo, P. Maddalena,E. Massera, E. Prudnikova, and E.
Santamato, J. Chem. Phys {\bf 107}, 9783 (1997).


\end{thebibliography}
\end{document}